\documentclass[12pt,preprint]{aastex}

\shorttitle{Changes in p-mode damping and excitation}
\shortauthors{Salabert and Jim\'enez-Reyes}

\begin{document}

\title{Damping and excitation variations of the solar acoustic modes \\ 
using LOWL observations}

\author{D. Salabert\altaffilmark{*}}
\affil{High Altitude Observatory, National Center for Atmospheric Research, P.O. Box 3000, Boulder, CO 80307-3000, USA}
\altaffiltext{*}{Now at: National Solar Observatory, 950 N. Cherry Avenue, Tucson AZ 85719, USA, dsalabert@nso.edu}

\and

\author{S.J. Jim\'enez-Reyes}
\affil{Instituto de Astrof\'isica de Canarias, 38205 La Laguna, Tenerife, Spain}

\begin{abstract}
We have used observations made with the helioseismic instrument LOWL
collected over $\sim$~6 years to carry out an independent study of 
the variations of the p-mode damping and excitation rates with solar activity.
We observe significant variations in the mode height, mode width and mode velocity power 
over a wide range of angular degree values. Their sensitivities to solar
activity show clear evidence of frequency dependence, the modes in the frequency range from 2700 and 3300~$\mu$Hz showing 
the largest variations and exhibiting a maximum change centered around 3100~$\mu$Hz.
As for the mode energy supply rate, it is consistent, at the level of precision of the observations,
with a zero change along the solar cycle and over the range of studied frequencies.
Moreover, the variations with solar activity of each of these parameters are observed to be more or less $\ell$-independent
over the range of studied angular degrees. Our results provide
the first in-depth confirmation of the findings obtained from GONG measurements
for intermediate angular degrees.

\end{abstract}


\section{INTRODUCTION}
Our knowledge and understanding of the Sun has improved considerably with the study
of the solar oscillations, leading to important modifications in the solar- and stellar-evolution models.
Indeed, through the accurate measurements of the frequencies of acoustic modes, helioseismology 
has represented a revolution in solar physics, revealing a detailed picture 
of the structure of the Sun and its internal rotation for instance \citep[see][for a detailed review]{thompson03}.
 Furthermore, in the last decade, the solar p-mode parameters 
have also been demonstrated to be very sensitive to changes in solar activity, 
showing that they can also help us
reach a better understanding of the processes driving the solar 
magnetic cycle, and of the physics of the acoustic modes themselves as, for example, the processes 
responsible for their excitation and damping.

The first reports about the variability of the p-mode parameters concerned
the mode frequencies. \citet{woodard85} found that modes with
degrees $\ell$=0 and $\ell$=1 presented a change in the central frequency of 0.42~$\pm$~0.14~$\mu$Hz.
Few years later, \citet{duvall88} and \citet{libbrecht90} also observed significant variations 
in the p-mode central frequency for intermediate degrees. The quantity of observations
increasing, thanks to ground-based networks and later on space-based instruments, deeper
analysis were carried out for low-degree modes \citep[see, e.g.,][]{chaplin98,chano98,salabert02,gelly02}
and intermediate and high-degree modes \citep[see, e.g.,][]{howe99,chano01b}, 
establishing, for instance, a strong $\ell$-dependence in the frequency shifts.
The study of the relations between the p-mode heights and widths with the solar cycle
is a more difficult task to achieve. Indeed, since their estimations are subject
to several observational effects and that long datasets are needed, the first reports of 
their variabilities with solar activity were sometimes contradictory. For example, \citet{palle90a} 
and \citet{jefferies90} reported an increase in mode widths with increasing activity in low- and medium-$\ell$ 
observations respectively, while no significant change were found by \citet{elsworth93} with full-disk data.  
Today, the quality and the amount of available data (some of the instruments are collecting data since 
more than a solar cycle) are such that it is now possible to have reliable estimates of the mode heights 
and widths and to study the variability with the solar cycle of the underlying p-mode damping and excitation.
Recent work achieved with low-$\ell$ data \citep[see, e.g.,][]{chaplin00,salabert03,chano03}
and also with intermediate-$\ell$ data \citep[see, e.g.,][]{komm00a,howe04} have shown significant 
temporal changes in the mode heights, mode widths, and mode energies 
(proportional to height$\times$width), as previously observed by \citet{palle90a,palle90b}, 
\citet{jefferies90}, and \citet{elsworth93}. 
All these recent studies also reported that the temporal variation of the p-mode excitation 
(proportional to height$\times$width$^2$) is consistent with a zero change.

Here, we made use of the spatially-resolved LOWL observations obtained
between beginning of 1994 and end of 1999 to carry out an independent
study of the temporal variations of the p-mode damping and excitation rates with  solar activity. 
By use of our own data analysis and peak-finding codes, 
we compared the observed changes for modes from $\ell=0$ to $\ell=99$ to the unique observations made
so far at intermediate-$\ell$ obtained with GONG data \citep{komm00a,komm00b}. 

\section{DATA AND ANALYSIS}
\subsection{Observations}
The single-site helioseismic instrument LOWL, located at the 
Mauna Loa Solar Observatory, Hawaii (USA), has been acquiring
data since February 1994. This instrument allows simultaneous
observations of low- and intermediate-degree solar oscillations
with spatial resolution. 
The LOWL instrument is a Doppler imager based on a Magneto-Optical
Filter. It employs a two-beam 
technique to simultaneously observe the solar disk in opposite wings 
of the absorption line of potassium at 769.9~nm. Images are obtained 
every 15~$\sec$ with a spatial resolution of
25~$\arcsec$. The intensity 
difference between
images obtained in both wings of the spectral line are related to the 
Doppler shift of the line, providing a measurement of the line-of-sight 
velocity of the observed area. The instrument is extremely stable 
against zero-point drifts and immune to noise source due to intensity 
fluctuations and image motions. Some of the main advantages of the 
instrument are its wide velocity range where the Doppler analyzer 
is linear, the absence of moving parts in the optics and
an improved thermal control system. These attributes render it  
one of the best experiments for the observations of low and 
intermediate-degree oscillations. A more detailed description of 
the instrument as well as a deeper discussion of its characteristics 
can be found in \citet{tomczyk95}.

The dopplergram (or velocity) images are obtained using the blue and red intensity
images obtained through the two transmission bands, and were averaged over 1 minute intervals. 
Once the velocity images calibrated,
a spherical harmonic decomposition is performed on each of them in order to create timeseries
for mode degrees from $\ell=0$ to $\ell=99$. A spectral analysis
of each of the timeseries allow us to extract the main features
of the acoustic modes. Detailed explanations of the LOWL data reduction
can be found in \citet{chano01a}.

The observations used in the present analysis begin March 1st 1994
and end October 12 1999, spanning
the decreasing phase and the minimum of the solar cycle 22 and the rising phase of
the solar cycle 23. They were divided into 19 independent
timeseries of 108 days with a 60~s cadence (Table~\ref{table:tseries}),
following the 36-d GONG month reference \citep[see, e.g.,][Table 1]{howe99}.
The duty cycles (or temporal fills) of the 
timeseries are very similar, about 20$\%$, and
since the observations were
obtained from just one site, the timeseries 
present regular gaps every 24 hours which induce temporal 
sidelobes, located at $\pm11.57\mu$Hz away from the
main peak. These sidelobes were taken into account
during the fitting process of
the p-mode parameters (Sec.~\ref{sec:fittingproc}). Note also that
the spectra with the 3 lowest fill values were removed for this analysis, 
so 16 108-d independent timeseries were used.
\clearpage
\begin{deluxetable}{ccccrrr}
\tabletypesize{\footnotesize}
\tablecolumns{7}
\tablewidth{0pt}
\tablecaption{Details of the 108-d LOWL timeseries
\label{table:tseries}}
\tablehead{
\colhead{Series \#} & \colhead{Start Date} & \colhead{End Date} & \colhead{Fill (\%)} 
& \colhead{$F_{10.7}$\tablenotemark{a}} & \colhead{$R_I$\tablenotemark{b}} 
& \colhead{KPMI\tablenotemark{c}} }
\startdata
\phn1   & 940301 & 940616 & 20 & 83.24  & 23.33  & 8.62  \\
\phn\phd2$^*$   & 940617 & 941002 & 16 & 80.01  & 27.08  & 8.20  \\
\phn3   & 941003 & 950118 & 23 & 81.44  & 26.92  & 8.77  \\
\phn4   & 950119 & 959596 & 22 & 82.07  & 25.76  & 8.44  \\
\phn5   & 950507 & 950822 & 23 & 76.85  & 14.48  & 7.99  \\
\phn6   & 950823 & 951208 & 18 & 74.50  & 15.73  & 7.42  \\
\phn7   & 951209 & 960325 & 22 & 70.49  & 8.37   & 7.06  \\
\phn8   & 960326 & 960711 & 26 & 71.61  & 8.20   & 6.60  \\
\phn9   & 960712 & 961027 & 20 & 71.32  & 5.82   & 6.40  \\
\phn10$^*$      & 961028 & 970212 & 10 & 74.19  & 11.97  & 7.19  \\
\phn11$^*$      & 970213 & 970531 & 16 & 74.19  & 12.55  & 6.83  \\
12      & 970601 & 970916 & 21 & 80.26  & 23.82  & 7.68  \\
13      & 970917 & 980102 & 18 & 92.14  & 34.81  & 9.22  \\
14      & 980103 & 980420 & 30 & 100.26 & 47.08  & 9.75  \\
15      & 980421 & 980806 & 24 & 111.70 & 61.40  & 11.65 \\
16      & 980807 & 981122 & 18 & 132.51 & 78.01  & 13.43 \\
17      & 981123 & 990310 & 20 & 140.01 & 72.10  & 14.15 \\
18      & 990311 & 990626 & 23 & 141.61 & 93.28  & 13.90 \\
19      & 990627 & 991012 & 21 & 161.31 & 97.41  & 15.33 \\
\enddata
\tablenotetext{a}{Integrated Solar Radio Flux at 10.7-cm, in $10^{-22}{\rm W~m}^{-2}{\rm Hz}^{-1}$}
\tablenotetext{b}{Sunspot Number, dimensionless}
\tablenotetext{c}{Kitt Peak Magnetic Index, in G}
\tablecomments{The timeseries denoted by an asterix ($^*$) were removed for this analysis.}
\end{deluxetable}
\clearpage

\subsection{Mode Parameter Estimation}
\label{sec:fittingproc}

LOWL is an instrument with imaging capability that makes the
study of a large range of solar p modes, from the low to intermediate 
acoustic mode degrees, possible. Due to the spherical symmetry of the Sun, the 
spherical harmonics are the most common of the spatial filters to isolate 
the information of each mode. However, the spherical
harmonics are not orthogonal over the observed area, limited to half
a sphere introducing correlations between different ($\ell$,$m$) 
modes. Therefore, the observed power spectra, in the case of resolved 
observations, are a linear combination of different modes, commonly 
represented by the so-called leakage matrix $C_{m,m'}^{\ell,\ell'}$. 
It describes the imperfect isolation of the individual modes. We will 
discern two types of leakage signal: the spatial leaks, from same 
($n$,$\ell$) modes, and the mode contamination from different 
($n$,$\ell$) modes. 

In this particular case, it can be proven that the statistics of the 
real and imaginary part of the Fourier transform of the ($\ell$,$m$) 
modes follow a multi-normal distribution described by the so-called
covariance matrix \citep{schou92,appour98}. However, as
the degree of the target mode increases, the covariance matrix becomes
nearly singular leading to a subsequent failure of the fitting procedure. 
We decided to perform a maximum-likelihood minimization using the
diagonal elements of the covariance matrix to estimate the set of 
parameters, $\vec{a}_{n,\ell}$, describing the p-mode profiles, it is:
\begin{equation}
  \label{eq:likelihood}
  S(\vec{a}_{n,\ell})= \sum_{m=-\ell}^{\ell} \sum_{i=1}^{N}\left[\ln
    {\mathcal{M}}_{n,\ell,m}(\nu_{i})  + 
    \frac{Y_{n,\ell,m}(\nu_{i})}{\mathcal{M}_{n,\ell,m}(\nu_i)}
    \right].
\end{equation}
The sum runs over all the points in frequency and over the (2$\ell$+1) 
$m$-components of the multiplet. The power spectrum of 
the target mode is denoted by $Y_{n,\ell,m}$, while $\mathcal{M}_{n,\ell,m}$ models the power spectrum and may be
written as the superposition of the spatial leaks and mode contamination,
i.e.: 
\begin{eqnarray}
  \mathcal{M}_{n,\ell,m}(\nu_{i}) & = &\sum_{m=-\ell}^{\ell} 
  H_{n,\ell} C_{m,m'}^{\ell,\ell} v_{n,\ell, m}(\nu) + \\
  & & \sum_{n',\ell',m'} H_{n',\ell'} C_{m,m'}^{\ell,\ell'} v_{n',\ell', m'}(\nu) + N_{n,\ell}, \nonumber 
\end{eqnarray}
where $v_{n,\ell,m}(\nu)$ is a non-asymmetric Lorenztian 
profile defined by: 
\begin{equation}
  \label{eq:lorentz}
  v_{n,\ell, m}(\nu)= \sum_{k=-3}^{3} \beta_{|k|} 
  \frac{\displaystyle H_{n,\ell} \  (\Gamma_{n,\ell}/2)^2}
       {\displaystyle (\nu - \nu_{n,\ell,m} + k_D \times k)^2 + (\Gamma_{n,\ell}/2)^2}.
\end{equation}
Thus, $H_{n,\ell}$, $\Gamma_{n,\ell}$, $N_{n,\ell}$
are respectively the mode height, the mode width and the background
noise, while $\nu_{n,\ell,m}$ is the central frequency 
for each $m$-component of the multiplet. $\beta_{|k|}$ is the ratio of 
the height of the sidelobes to the height of the main peak, estimated 
for each time series and $k_D$ is the constant separation of the 
sidelobes, equal to 11.57$\mu$Hz. Note we have taken into account the 
3 first sidelobes. As observed, for example, by \citet{duvall93}, 
the p-mode profiles possess small but significant levels of asymmetry:
however in the case of the present analysis, the signal-to-noise ratio obtained with observations from one single-site 
instrument does not allow reliable estimations of the asymmetry parameter. In order to constrain the number
of free parameters during the minimization process, we assumed that each $m$-component is well
described by a non-asymmetric Lorenztian profile (eq.~\ref{eq:lorentz}). 

Regarding the central frequency for each $m$-component $\nu_{n,\ell,m}$,
it is represented by: 
\begin{equation}
  \label{eq:nupoly}
  \nu_{n,\ell,m}=\nu_{n,\ell} + \sum_{i=1}^{n} a_i(n,\ell) \mathcal{P}_i^{\ell}(m).
\end{equation}
The polynomials $\mathcal{P}_{i}^{\ell}(m)$ are the \citet{ritz91} polynomials 
defined in \citet[][App.~A]{schou94}. The coefficients $a_i(n,\ell)$ 
represent the shift in frequency induced mainly by the internal 
rotation. The number of $a$-coefficients was fixed to 1 for $\ell=1$; 
4 for $\ell=2$ and $\ell=3$; 6 for $4\leq\ell\leq9$ and 9 for 
$\ell\geq10$. 

The mode parameters are constrained by performing simultaneous 
fits of the spatial leaks with the implicit assumption that the mode
contamination is known. The process becomes thus iterative, improving
in each iteration the mode contamination dominated (for intermediate
degrees) by $n-n'=\pm1$ and $\ell'-\ell=\pm3$. The process is iterated
until the mode frequency changes between iterations drops below
a given threshold. Finally, the fits are performed inside a narrow 
window centered on each $m$-component. 

Using initial guesses obtained  from a collapsed diagram \citep{chano01a}, 
an iterative process was used to converge to the best estimates of 
the mode parameters of the 16 108-d independent series. The natural 
logarithms of the mode height, width, and background noise have 
been fitted, not the parameters themselves, resulting in a normal 
distribution which allows the uncertainties on each parameter to be 
determined from the inverse Hessian matrix.



\subsection{Formulation of the Acoustic Spectrum}
The mode damping rate, $\eta_{n,\ell}$, is directly related to the mode width, $\Gamma_{n,\ell}$, by:

  \begin{equation}
  \label{eq:damping}
  \eta_{n,\ell}=\pi\Gamma_{n,\ell}.
  \end{equation}
The velocity power, or mode power, $P_{n,\ell}$, of a given mode $(n,\ell)$ 
corresponds to the area under the mode, which is a combination of the
mode height, $H_{n,\ell}$, and the mode width, $\Gamma_{n,\ell}$. It can be written as:

  \begin{equation}
  \label{eq:velpower}
  P_{n,\ell}=\frac{\pi}{2}C_{\rm obs}H_{n,\ell}\Gamma_{n,\ell},
  \end{equation}
where $C_{\rm obs}$ is a constant to correct for the effects of
the observational techniques.
Using the analogy of the harmonic damped oscillator, the energy supplied 
to the modes, or energy supply rate, is estimated by:

  \begin{equation}
  \label{eq:esupply}
  \dot{E}_{n,\ell}=\frac{dE_{n,\ell}}{dt}=2\pi E_{n,\ell}\Gamma_{n,\ell},
  \end{equation}
where $E_{n\ell}$ is the total mode energy \citep{goldreich94}, which is given by:

  \begin{equation}
  \label{eq:energy}
  E_{n\ell}=M_{n,\ell}P_{n,\ell}.
  \end{equation}
$M_{n,\ell}$ corresponds to the mode masses, which are inversely proportional to the mode inertia, $I_{n,\ell}$ \citep[see][]{jdc96}. 

In what follows, we focus on studying the temporal variations in the damping and the excitation rates of the p-mode oscillations. Since the peak-finding procedure returns the best estimates in natural logarithm values of the mode height and width, any temporal variations in these parameters (and any additive combinations of the two) will correspond to the fractional variations of these parameters (and any multiplicative combinations of the two). The constant $C_{\rm obs}$ and the mode masses, $M_{n,\ell}$, are assumed to remain constant over the solar cycle, and then can be ignored in the present analysis. Concerning the determination of the uncertainties
in the variations of the mode velocity power, $P_{n,\ell}$, and the energy supply rate, $\dot{E}_{n,\ell}$,
they are estimated by the formula for non-independent variables, taking
into account the strong anticorrelation (taken to be equal to -0.9) between the fitted mode width, $\Gamma_{n,\ell}$,
and height, $H_{n,\ell}$ \citep[see, e.g.,][]{chaplin00}. 

Hereafter, we represent the logarithm values of $H_{n,\ell}$, $\Gamma_{n,\ell}$, $P_{n,\ell}$, and $\dot{E}_{n,\ell}$  by $h_{n,\ell}$, $\gamma_{n,\ell}$, $p_{n,\ell}$, and $\dot{e}_{n,\ell}$ respectively.
\clearpage
\begin{figure}[t]
\centering
\epsscale{1}
\plotone{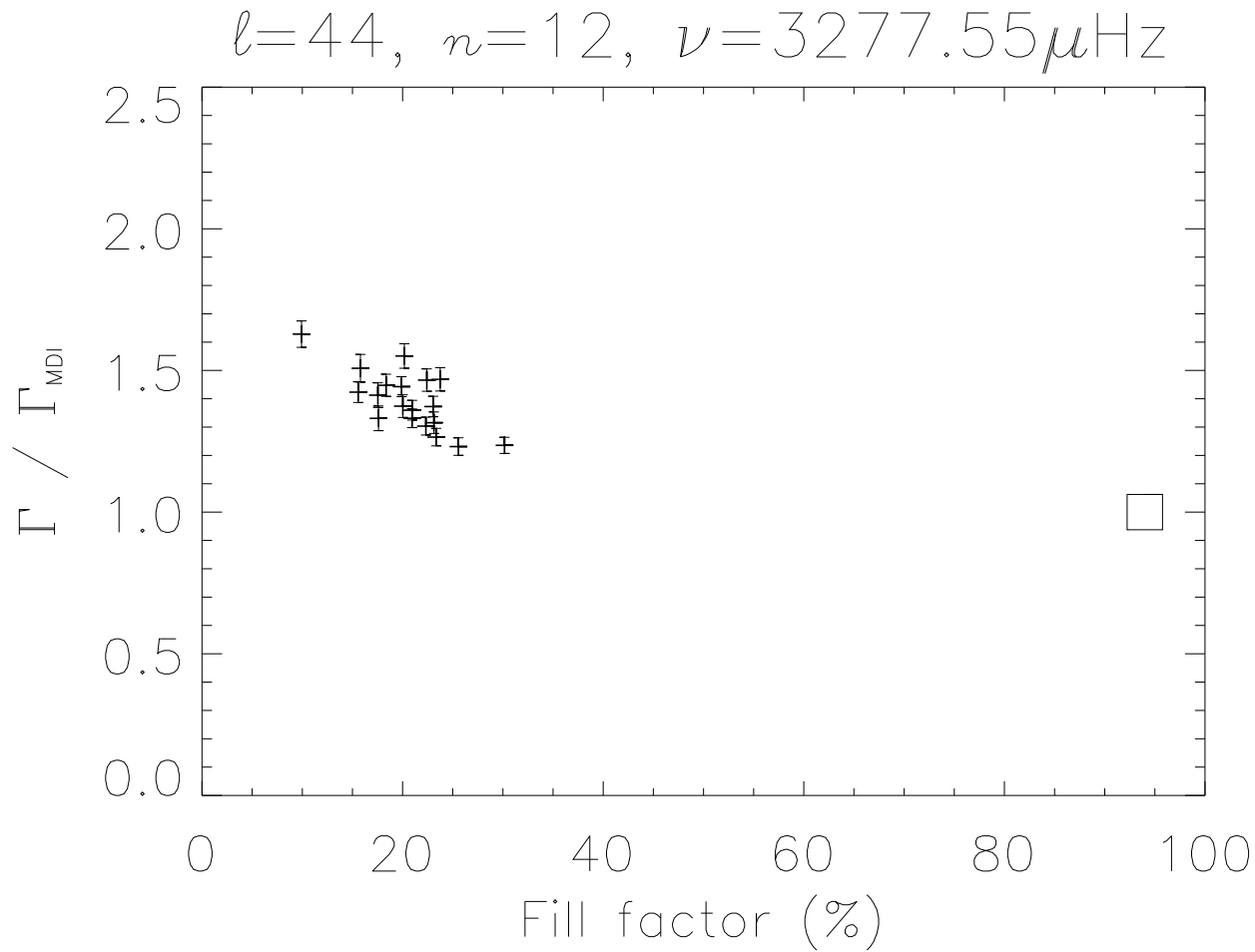}
\caption{Example of the variation of the mode width, $\Gamma$, in the case of the mode $n=12$, $\ell=44$, 
as a function of the fill factor. The LOWL mode width estimates (+) were normalized by the corresponding SOHO/MDI value ($\square$).}
\label{fig:width_mdi}
\end{figure}

\begin{figure*}[t]
\centering
{\includegraphics[scale=0.47]{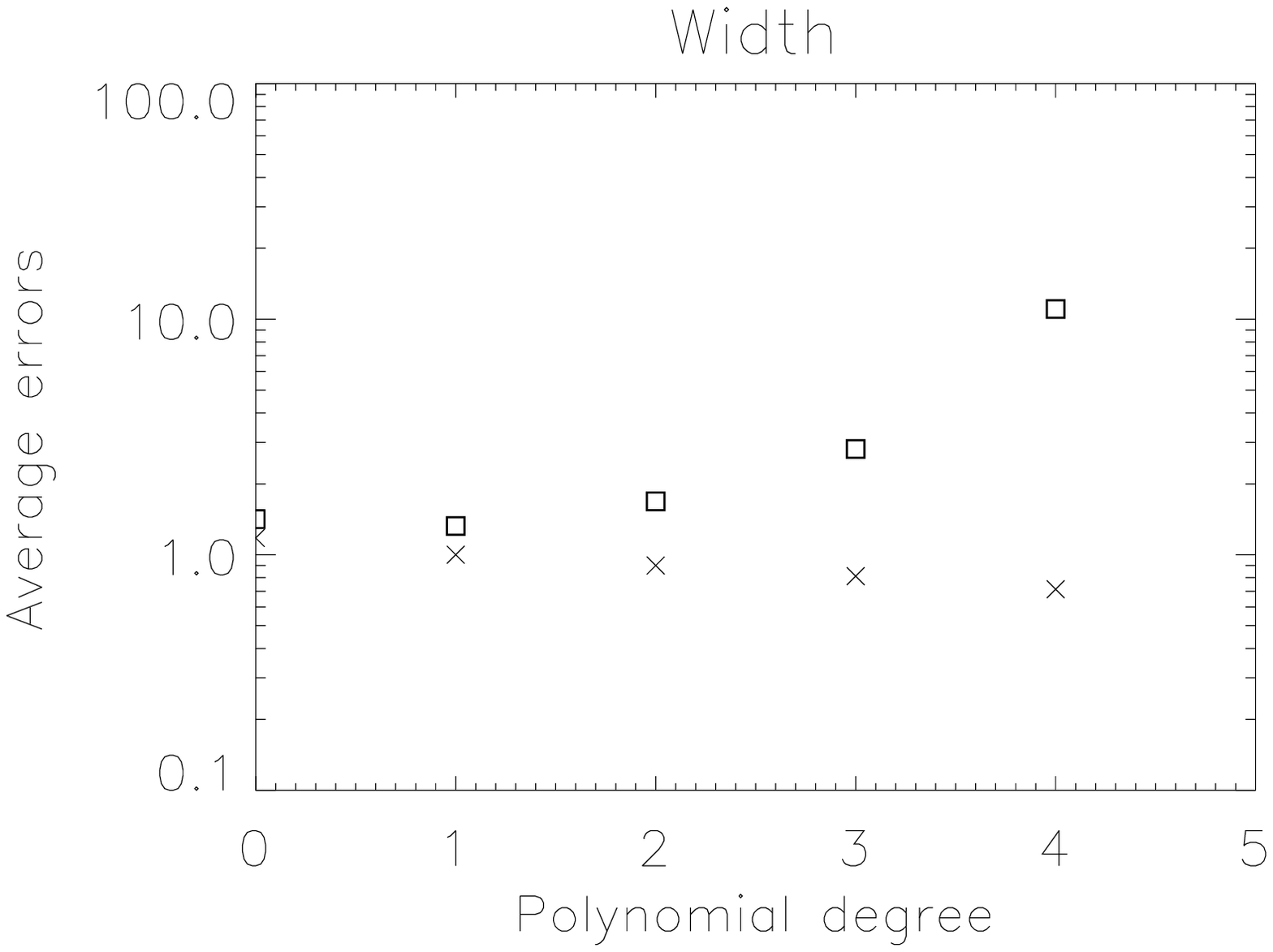}
 \includegraphics[scale=0.47]{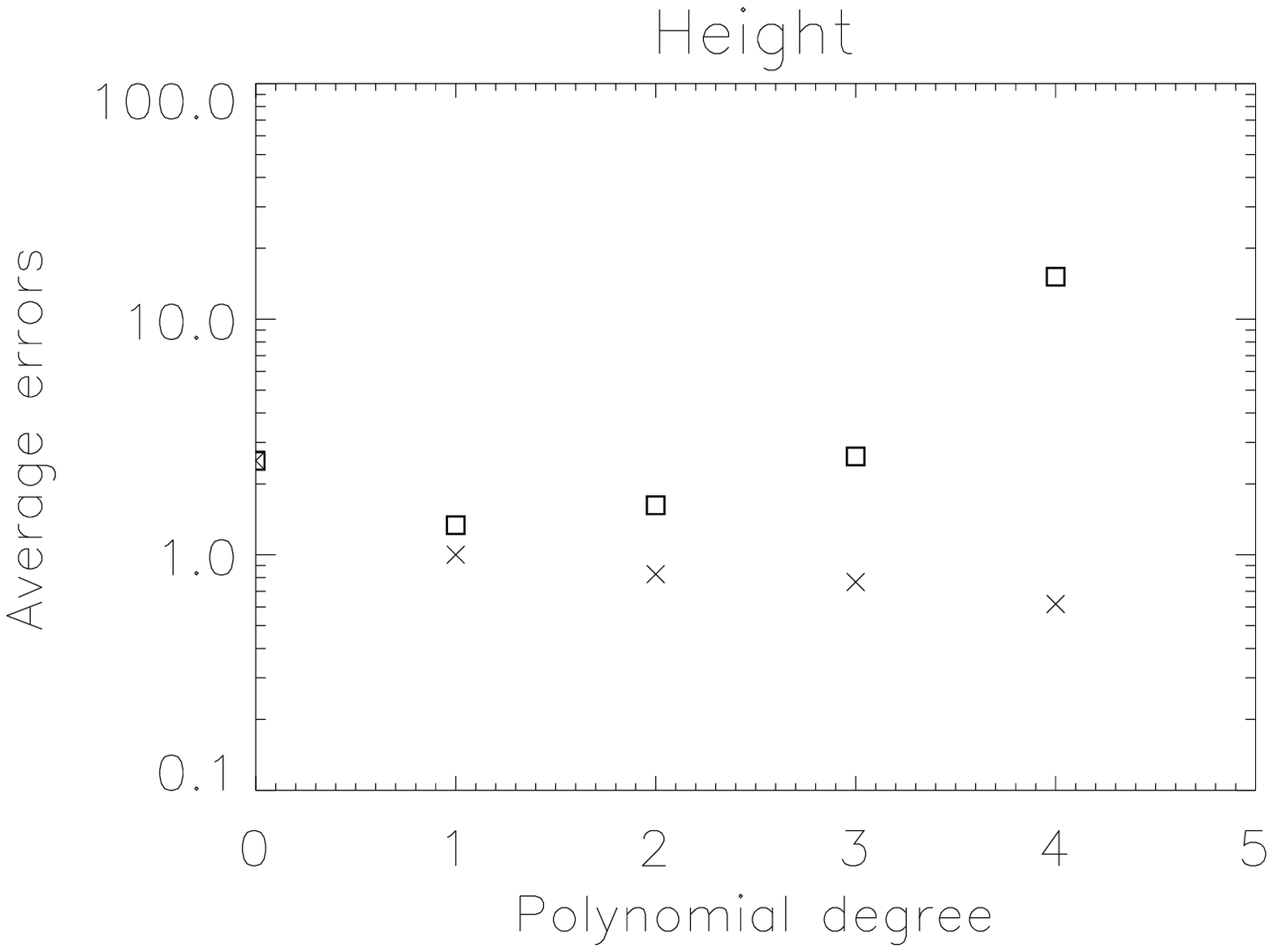}}
\caption{Average in-sample errors ($\times$) and out-sample errors ($\square$) 
of polynomial fits of the mode widths, $\Gamma$ ({\it left panel}) 
and of the mode heights, 
$H$ ({\it right panel}) as a function of the fill factor for different degrees 
of polynomial.}
\label{fig:in-out_error}
\end{figure*}
\clearpage

\subsection{Temporal Window Correction}
\label{sec:wcorr}
The presence of gaps in the observations leads to a redistribution of power 
from the main lobe into sidelobes and into the background. The estimates of the 
mode heights and of the mode widths will be then affected, the heights
being underestimated and the widths overestimated. As shown in \citet{komm00a}, the observed bias
is a function of the mode degree and the mode frequency.
Moreover, the effects due to the presence of gaps will be larger as the fill factor decreases.

Figure~\ref{fig:width_mdi} shows an example of the variation in the estimation of the
mode width (for $n$=12, $\ell=44$) between a fill of nearly 100\% and the actual LOWL fill factors. 
The plotted LOWL width estimates were normalized by the corresponding value obtained
with contemporaneous timeseries from the space-based instrument SOHO/MDI, which is supposedly 
very close to the {\it true} value. \citet{komm00a}, using GONG data, showed that a simple 
linear model provides an adequate description of the overestimation of the mode widths
as the fill factor decreases\footnote{\citet{komm00a} also applied equation~\ref{eq:fillcorr_komm} to correct the observed
bias from the fill factor on the other mode parameters.}, as:
 
  \begin{equation}
  \Gamma_i^{\rm{D}}=c_0+c_{\rm{D}}\times(1-{\rm{D}}),
  \label{eq:fillcorr_komm}
  \end{equation} 
where D is the value of the fractional temporal fill; $c_0$ and $c_{\rm{D}}$ are the intercept 
and the gradient respectively. In that simple model, the intercept $c_0$ corresponds
to the extrapolated width value for 100\% fill, which supposes that this model
is correct up to a fill of 100\%. Even though this linear description is correct over
 a limited range 
of fills close to an ideal fill of 100\%, it is not the case over a
large range of duty cycles as we can see in Figure~\ref{fig:width_mdi}.

In this paper, we are interested in the temporal variations of the damping and 
excitation p-mode parameters, 
and not in their absolute estimates. Then, we can modify equation~\ref{eq:fillcorr_komm} in
order to have a simple model which describes the bias from the fill factor and corrects it
to the mean fill value, $\langle\rm{D}\rangle$, of the analysed LOWL timeseries, instead of
a 100\% fill. 

We also checked the degree of 
the polynomial necessary to describe such bias. We followed the cross-validation method described
in \citet{komm00a} and adapted from \citet{gershenfeld99} and \citet{goutte97}. It consists in dividing randomly our data sets into two subsets, fit one 
subset with the 
polynomial and calculate the average errors between the model and the fitted subset 
(also called 
{\it in-sample error}), and the average errors between the model and the excluded subset 
(or {\it out-sample error}),
and repeat this procedure for several polynomial degrees. 
The in-sample error is expected to decrease with increasing model
complexity, while the out-sample error will decrease up to a certain degree and then 
increase again
 when the model overfits the data. The minimum in the out-sample error as a function of 
the polynomial
degree defines the best model.

The left panel of Figure~\ref{fig:in-out_error} shows the corresponding results for 
the mode widths, and
the right panel for the mode heights. The original 16 timeseries were divided into two 
subsets of 8 series each. On Figure~\ref{fig:in-out_error}, the average in- and out-sample 
errors were normalized by the 
in-sample error of the polynomial of degree 1. On both panels, the in-sample errors 
(crosses) decrease
with an increasing polynomial degree, whereas the out-sample errors (squares) shows 
a minimum 
for a degree of polynomial fit equal to 1, indicating that a linear fit is the best model 
describing the dependence of the mode widths
and heights with the fill factor. 

Therefore, we decided to describe by a linear polynomial the bias introduced by 
the window functions on the p-mode parameters at each ($n,\ell$) multiplet, 
but using the mean fill value, $\langle\rm{D}\rangle$, instead of a 100\% fill, as:

  \begin{equation}
  Y_i^{\rm{D}}=c_0+c_{\rm{D}}\times({\langle\rm{D}\rangle-\rm{D}}),
  \label{eq:fillcorr}
  \end{equation}  
where $Y_i^{\rm D}$ stands for any of the mode parameters, where 
$i\equiv[h_{n,\ell},\gamma_{n,\ell},p_{n,\ell},\dot{e}_{n,\ell}]$, and D is the fill factor. 
$c_0$ and $c_{\rm{D}}$ are again the intercept and the gradient respectively.
When looking for temporal variations in the p-mode parameters, equation~\ref{eq:fillcorr} 
is a sufficient description of the temporal window effects over a limited 
range of fill factor values. Ideally, by modulating the SOHO/MDI data with a large range 
of window functions between 
a 100\% fill  and fills around 20\% (typical fill factor for a single-site instrument, as LOWL), 
we could determine an emperical model to describe the non-linear dependence  between the 
damping and excitation p-mode parameters and the fill factors in a range from  
a 100\% fill to lower duty cycles. This issue will be addressed in a future work.

\subsection{Solar-Cycle Variations}
To search for solar-cycle variations, we add to equation~\ref{eq:fillcorr} a polynomial term function 
of a solar activity proxy (SAP). Then equation~\ref{eq:fillcorr} becomes:

  \begin{equation}
  \label{eq:lincorr}
   Y_i^{\rm{D,SAP}}=c_0+c_{\rm{D}}\times({\langle{\rm D}\rangle}-\rm{D})+\delta Y_i^{\rm{SAP}}\times\rm{SAP},
  \end{equation}
 where $\delta Y_i^{\rm{SAP}}$ is a measure of the fractional changes of each of the 
parameters $i$ per unit of solar activity. The linear dependence between the p-mode 
parameters
and solar activity has already been demonstrated by previous studies using low-degree 
observations 
\citep[see, e.g.,][]{chaplin00,salabert03,chano03} and using intermediate-degree
observations \citep[see, e.g.,][]{komm00a}. By use of the multi-linear regression 
described by equation~\ref{eq:lincorr}, we extracted 
the fractional temporal changes $\delta h_{n,\ell}$, $\delta\gamma_{n,\ell}$, $\delta p_{n,\ell}$
 and $\delta\dot{e}_{n,\ell}$, corresponding respectively to the changes in the mode 
height, width, velocity power and energy supply rate, corrected from the bias due to
the temporal fill. This regression was performed on the estimates of 16 108-d independent 
timeseries, weighted by the uncertainties returned by the peak-finding procedure, 
and for each mode frequency and degree. 
Over all the studied frequency range and mode degrees, the changes in mode parameters 
due to solar activity are about 3 orders of magnitude smaller than the corresponding changes due to the fill factor. 
\citet{komm00a} reported also larger effects due to the fill factor in GONG data in comparison to
the effects from solar activity, but of about 1 order of magnitude. The reason for this difference of order
of magnitude between LOWL and GONG observations is because the effects
of the fill factor are not linear in a range from high duty cycles to lower duty cycles.

We studied these temporal variations in p-mode parameters using 3 indices of global surface activity: 
the Kitt Peak Magnetic Index (KPMI), obtained from the Kitt Peak 
magnetograms\footnote{Data available at \texttt{http://nsokp.nso.edu/}}; the Sunspot Number ($R_I$), and 
the Integrated Radio Flux at 10.7~cm ($F_{10.7}$), the two last indices obtained from the National
Geophysical Data Center\footnote{Data available at \texttt{http://spidr.ngdc.noaa.gov/spidr/}}.  
A detailed discussion about the relations among several solar activity proxies can be found in \citet{bachmann94}.
\clearpage
\begin{figure}[t]
\centering
\resizebox{\hsize}{!}{\includegraphics{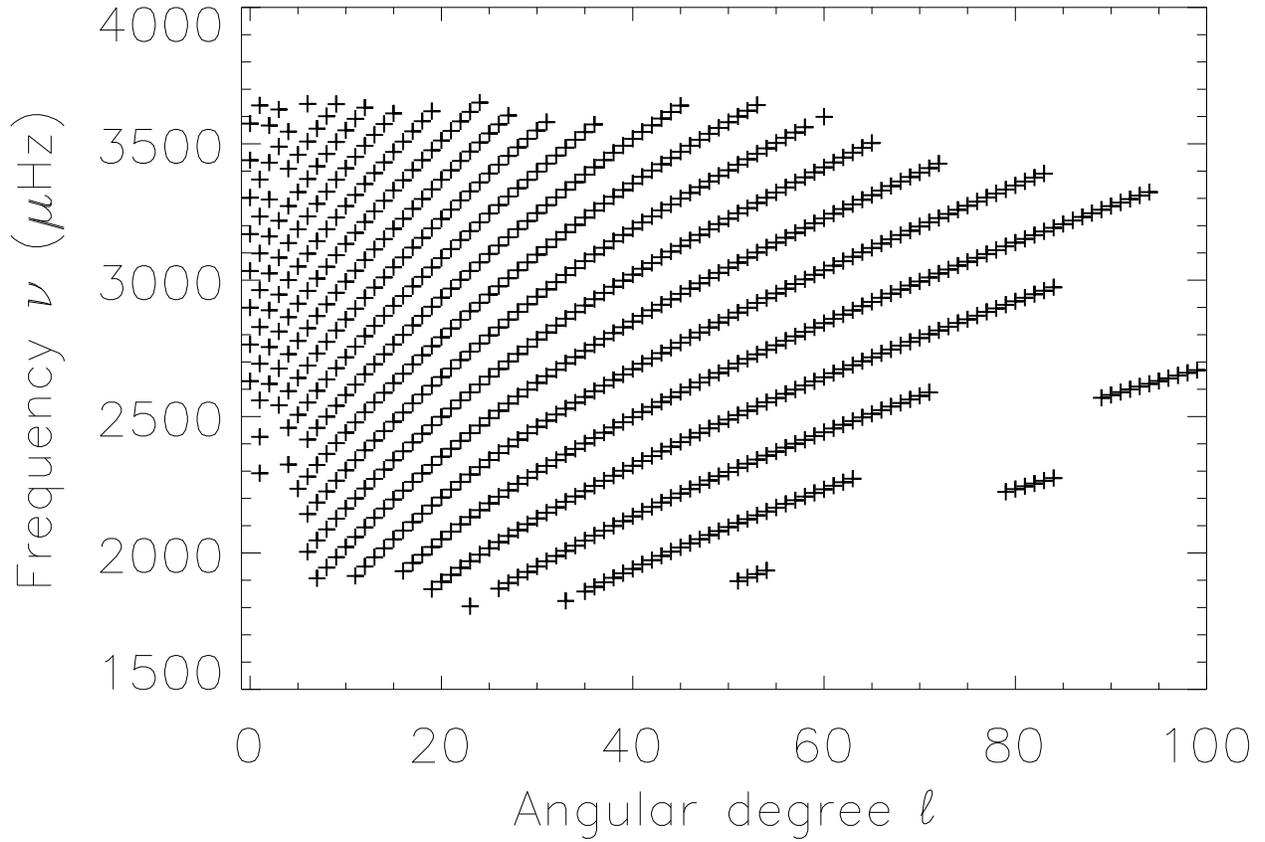}}
\caption{Set of common modes between the analyzed 108-d independent LOWL timeseries.
The large gap of missing modes from $\ell\sim60$ corresponds to the modes where
$(\delta\nu/\delta\ell -11.57) < 2\Gamma$.}
\label{fig:lnu}
\end{figure}

\begin{figure*}[t]
\epsscale{1}
\plotone{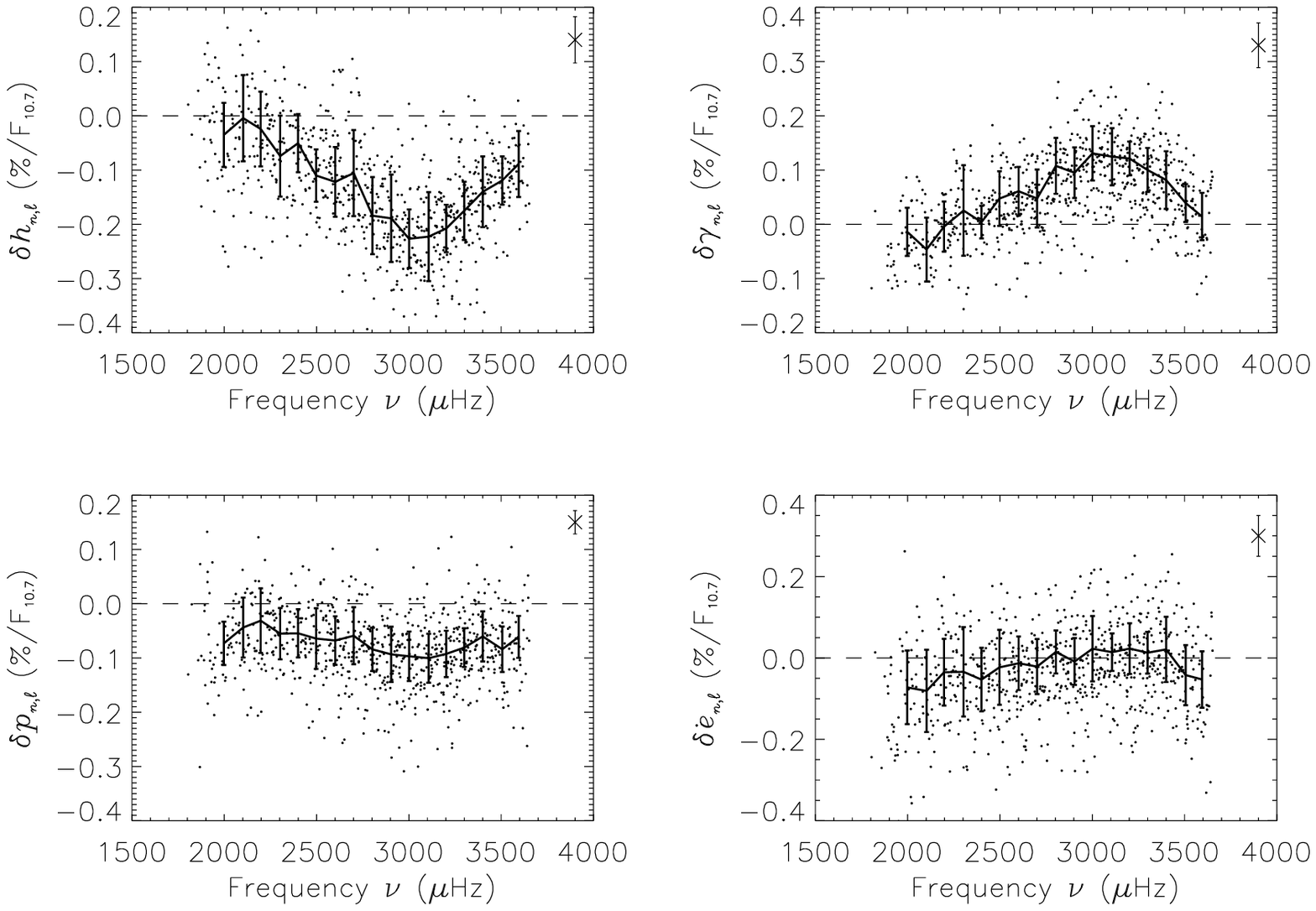}
\caption{Fractional variations (\% per unit of change in $F_{10.7}$) of the mode height, $\delta h_{n,\ell}$; the mode width, $\delta\gamma_{n,\ell}$; the mode velocity power, $\delta p_{n,\ell}$; and the mode energy supply rate, $\delta \dot{e}_{n,\ell}$, for $\ell=0$ to $\ell=99$ as a function of the frequency. The representative error bars in each upper-right panel correspond to the mean uncertainty of the fractional variations. The solid lines correspond to the averages over $30\leq\ell\leq60$ over frequency bins of 100 $\mu$Hz, with their corresponding $\pm$1 standard deviations. The dashed lines represent a zero change.}
\label{fig:fracnuflux}
\end{figure*}

\begin{figure*}[t]
\centering
\epsscale{1}
\plotone{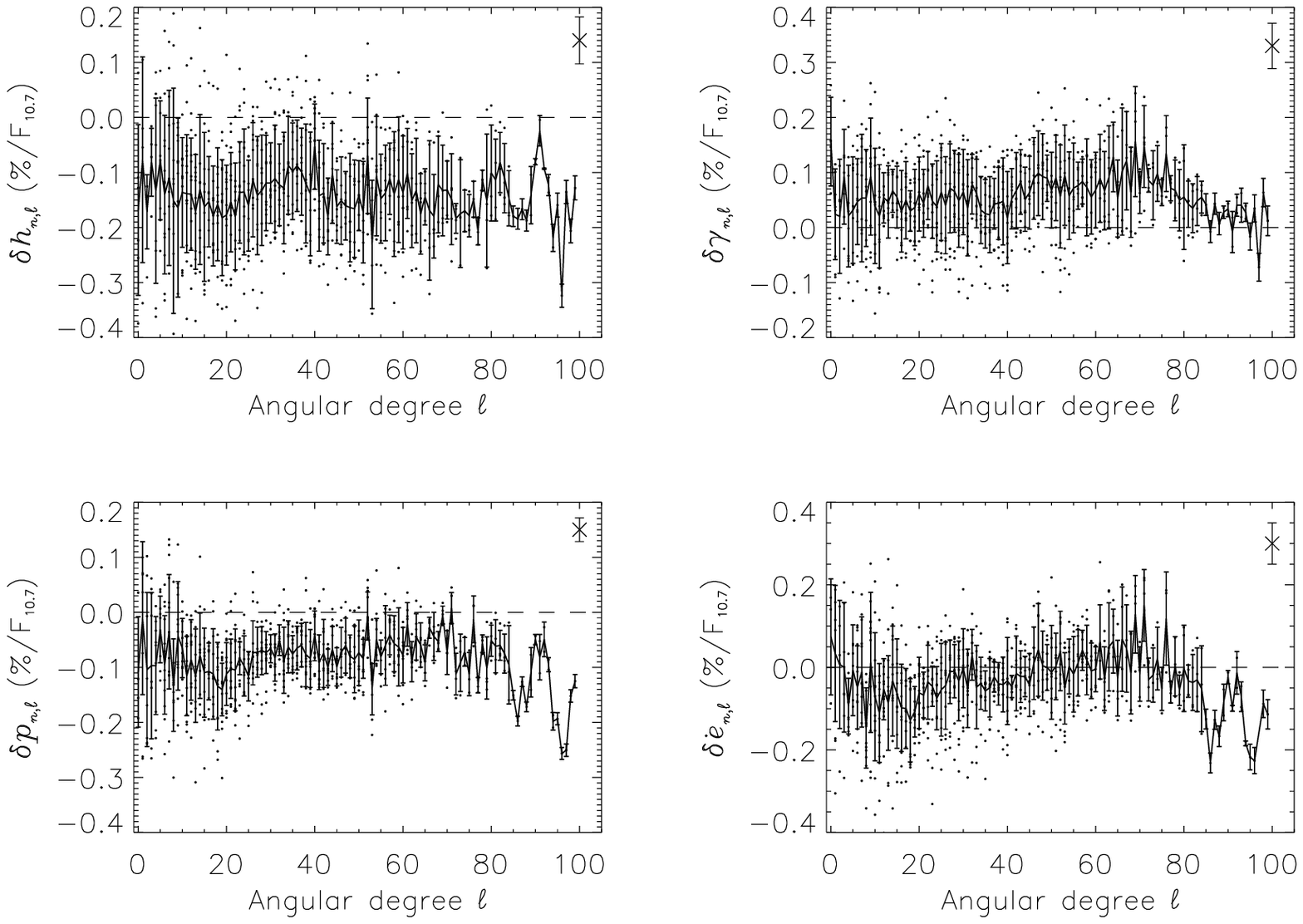}
\caption{Fractional variations (\% per unit of change in $F_{10.7}$) of the mode height, $\delta h_{n,\ell}$; the mode width, $\delta\gamma_{n,\ell}$; the mode velocity power, $\delta p_{n,\ell}$; and the mode energy supply rate, $\delta \dot{e}_{n,\ell}$, for $\ell=0$ to $\ell=99$ as a function of the mode degree. The representative error bars in each upper-right panel correspond to the mean uncertainty of the fractional variations. The solid lines correspond to the averages for each degree $\ell$ over the frequency range $2000\leq\nu\leq3500 \mu$Hz, with their corresponding $\pm$1 standard deviations. The dashed lines represent a zero change.}
\label{fig:fracdegflux}
\end{figure*}
\clearpage

\section{RESULTS}
\label{sec:results}
The multi-linear regression (eq.~\ref{eq:lincorr}) was performed on the good estimates defined
by a set of good-fit criteria in order to remove obvious unreasonable values, as: 
(1) the mode width must be within a factor of 5
of its initial guess; 
(2) the difference between the mode frequency and its initial guess 
must be less than the corresponding initial guess in mode width; and (3) the uncertainty of the 
mode frequency must be less than the initial guess in mode width. 
Moreover, in order to remove the possible systematic errors
due to the blending of the leaked modes and their sidelobes into the target modes, 
we only kept the modes with $(\delta\nu/\delta\ell -11.57) > 2\Gamma$.
About 20\% of the initial fitted modes were then removed from the estimate tables. 
The set of common modes used in the following analysis is composed of 771 modes (Figure~\ref{fig:lnu}).

Figure~\ref{fig:fracnuflux} shows the fractional changes in percent
of the studied mode parameters per unit of change in the solar radio 
flux, $F_{10.7}$, for the mode degrees $0\leq\ell\leq99$ as a function of the 
frequency. The solid lines correspond to the averages ($\pm$1 standard deviations) for modes
between $\ell=30$ and $\ell=60$ over frequency bins of 100 $\mu$Hz.
The degree dependence of the fractional changes of the mode parameters 
per unit of change in $F_{10.7}$ is shown in Figure~\ref{fig:fracdegflux}.
The solid lines correspond to the averages ($\pm$1 standard deviations) for each $\ell$ in
the frequency range between 2000 and 3500~$\mu$Hz. 
For modes with $\ell > 60$, the changes in mode height, power and energy supply rate
show large fluctuations, whereas 
the change in mode width show a slight decrease towards zero. These changes of 
behaviour as a function of $\ell$ are most probably not related
to solar activity, but rather could be the result of lower quality estimates of
the parameters due to the blending of the leaked modes with the 
target modes, and also because of a much lower number of estimates. For these
reasons, we limited all our discussions throughout this
paper for modes with $\ell \leq 60$.
The mean fractional changes times $10^3$ per unit of change in the 3 selected 
solar proxies (radio flux $F_{10.7}$; sunspot number $R_I$; and KPMI) 
are reported in Table~\ref{table:change} for several ranges of mode degrees and frequencies.
In order to extract global changes in percent of these parameters over the solar cycle, 
we computed also their average changes between $30\leq\ell\leq60$ and $2700\leq\nu\leq3300\mu$Hz, 
multiplied by the corresponding observed changes in the $F_{10.7}$, $R_I$, and KPMI indices
(see Table~\ref{table:global}).

The mode heights decrease 
with an increase in activity (upper-left panel on Figure~\ref{fig:fracnuflux}), 
with a fractional change $\delta h_{n,\ell}$ of $-1.30~\pm~1.01$ (times $10^3$ per unit of
 change in $F_{10.7}$) over all 
the frequency range and between $\ell=0$ and $\ell=60$. $\delta h_{n,\ell}$ shows 
the largest variation between 2700 and 3300~$\mu$Hz, with a change 
of $-2.01~\pm~0.92$ (times $10^3$ per $F_{10.7}$). 
For this range of modes, the p-mode heights change by about -25\% between the 
minimum and the maximum of the solar cycle (see Table~\ref{table:global}).
$\delta h_{n,\ell}$ presents a strong frequency dependence, with a 
maximum change centered around 3100~$\mu$Hz. Even though the mode height changes 
are negative for all the $\ell$ values, with almost no $\ell$-dependence, $\delta h_{n,\ell}$ 
decreases with decreasing $\ell$-value from $\ell\sim30$ up to $\ell\sim20$ 
followed by an increase for low-degree modes (upper-left panel on Figure~\ref{fig:fracdegflux}).
\citet{komm00a} also reported a decrease in this range of $\ell$-values in GONG data.

The mode widths increase with an increase in activity (upper-right panel on Figure~\ref{fig:fracnuflux}), 
 with a fractional change $\delta\gamma_{n,\ell}$ of $0.51~\pm~0.77$ (times $10^3$ per $F_{10.7}$). 
As $\delta h_{n,\ell}$, $\delta\gamma_{n,\ell}$ shows the largest departure from zero in the frequency range
from 2700 to 3300~$\mu$Hz, with an average change of $0.94~\pm~0.65$ 
(times $10^3$ per $F_{10.7}$), and  presents a maximum change 
centered around 3100~$\mu$Hz. Between the minimum and the maximum of the solar cycle,
the p-mode widths increase by about 14\%.
The mode width changes remain positive over all the range of 
studied degrees (upper-right panel on Figure~\ref{fig:fracdegflux}), and do not show an increase
with decreasing $\ell$-value as suggested by \citet{komm00a} with GONG observations.

The mode velocity powers exhibit a decrease 
with an increasing activity (lower-left panel on Figure~\ref{fig:fracnuflux}), 
with a change $\delta p_{n,\ell}$ of $-0.82~\pm~0.68$ (times $10^3$ per $F_{10.7}$) between $\ell=0$ 
and $\ell=60$ and over all the frequency range and a largest
decrease between 2700 and 3300~$\mu$Hz of $-1.03~\pm~0.71$ (times $10^3$ per $F_{10.7}$). 
The mode power decreases by an amount of 12\% over the solar cycle.
Even though the frequency dependence of $\delta p_{n,\ell}$ is less important than
for $\delta h_{n,\ell}$ and $\delta\gamma_{n,\ell}$, $\delta p_{n,\ell}$
presents also a maximum change around 3100~$\mu$Hz. 
The changes in mode power are negative
for all the studied $\ell$ values with no degree dependence for $\ell>30$ 
(lower-left panel on Figure~\ref{fig:fracdegflux}). 
However, also reported in \citet{komm00a}, below $\ell=30$, $\delta p_{n,\ell}$ presents
the same fluctuations as $\delta h_{n,\ell}$.

The mode energy supply rates
(lower-right panel on Figure~\ref{fig:fracnuflux}) present a mean change $\delta \dot{e}_{n,\ell}$ 
of $-0.41~\pm1.04$ (times $10^3$ per $F_{10.7}$) over all the frequency range and 
between $\ell=0$ and $\ell=60$,
and of $-0.21~\pm~0.94$ (times $10^3$ per $F_{10.7}$) between 2700 and 3300~$\mu$Hz. 
There might be some hints of a slight decrease with an increase in activity for frequencies below 3100~$\mu$Hz,
as already mentionned in \citet{komm00b}. However, at the level of precision of the data, the energy supply rate is
consistent with a zero change over all the studied frequency range along the solar cycle, in contrast
to the other parameters which are highly frequency dependent.
As a function of the mode degree, no $\ell$ dependence in $\delta \dot{e}_{n,\ell}$ can 
be observed between $\ell=30$ and $\ell=60$, and is consistent 
with a zero change (lower-right panel on Figure~\ref{fig:fracdegflux}).

Similar results, summarized in Tables~\ref{table:change} and \ref{table:global}, are obtained with 
the sunspot number $R_I$, and the KPMI index. 
Overall, $\delta h_{n,\ell}$, $\delta\gamma_{n,\ell}$, $\delta p_{n,\ell}$ and 
$\delta \dot{e}_{n,\ell}$ observed with LOWL data confirm the previous observations of variations in the p-mode damping and excitation rates made at intermediate degrees obtained with GONG data \citep{komm00a,komm00b}.

Even though this analysis went down as far as $\ell$=0, the precision in the observed changes
for low-degree modes ($\ell\leq3$) was such that the fractional variations were not returned at a level significant
enough to perform a detailed comparison with previous results obtained with full-disk observations, which also showed no variation of the energy supply rate along the solar cycle \citep[see, e.g.,][]{chaplin00,salabert03,chano03}. However, the low degree modes observed here show comparable amounts of changes than the intermediate degrees, at the level of precision of the data (Figure~\ref{fig:fracdegflux}). 
\clearpage
\begin{deluxetable}{rrrrrrrrrrrr} 
\tablecolumns{12} 
\tablewidth{0pt} 
\tabletypesize{\footnotesize}
\tablecaption{{Fractional changes $\times$ $10^3$ per unit change in solar activity proxy} 
\label{table:change}}
\tablehead{ 
\colhead{}    &  \multicolumn{3}{c}{$F_{10.7}$} &   \colhead{}   & 
\multicolumn{3}{c}{$R_I$} &   \colhead{}   & \multicolumn{3}{c}{KPMI}  \\ 
\cline{2-4} \cline{6-8} \cline{10-12}\\ 
\colhead{$\delta Y_i$} & 
\colhead{$\langle \delta Y_i \rangle \pm \sigma_{\langle \delta Y_i \rangle}$}   & 
\colhead{N} & \colhead{$\sigma$} & \colhead{} & 
\colhead{$\langle \delta Y_i \rangle \pm \sigma_{\langle \delta Y_i \rangle}$ }   &  \colhead{N} & 
\colhead{$\sigma$} & \colhead{} & 
\colhead{$\langle \delta Y_i \rangle \pm \sigma_{\langle \delta Y_i \rangle}$}   &  \colhead{N} & 
\colhead{$\sigma$}}
\startdata 
\cutinhead{$0\leq\ell\leq60$ and $1800\leq\nu\leq3600~\mu$Hz}

$\delta h_{n,\ell}$....... & -1.30~$\pm$~1.01 & 607 & -1.2 &  
 & -1.31~$\pm$~1.00 & 595 & -1.2 &
 & -16.12~$\pm$~10.57 & 596 & -1.4 \\ 

$\delta\gamma_{n,\ell}$....... & 0.51~$\pm$~0.77 & 586 & 1.0 & 
 & 0.52~$\pm$~0.75 & 587 & 1.0 & 
 & 7.04~$\pm$~8.20 & 594 & 1.0 \\ 

$\delta p_{n,\ell}$....... & -0.82~$\pm$~0.68 & 610 & -1.2 & 
 & -0.86~$\pm$~0.63 & 605 & -1.3 & 
 & -9.42~$\pm$~7.14& 613 & -1.3 \\ 

$\delta \dot{e}_{n,\ell}$....... & -0.41~$\pm$~1.04 & 606 & -0.4 & 
 & -0.41~$\pm$~0.97 & 598 & -0.5 & 
 & -3.10~$\pm$~10.80 & 593 & -0.3 \\ 

\cutinhead{$0\leq\ell\leq60$ and $2700\leq\nu\leq3300~\mu$Hz} 

$\delta h_{n,\ell}$....... & -2.01~$\pm$~0.92 & 225 & -1.6 & 
 & -2.00~$\pm$~0.86 & 223 & -1.6 & 
 & -23.37~$\pm$~9.14 & 224 & -1.8 \\ 

$\delta\gamma_{n,\ell}$....... & 0.94~$\pm$~0.65 & 227 & 1.2 & 
 & 0.94~$\pm$~0.61 & 227 & 1.2 & 
 & 11.73~$\pm$~6.72 & 228 & 1.5 \\ 

$\delta p_{n,\ell}$....... & -1.03~$\pm$~0.71 & 229 & -1.4 & 
 & -1.08~$\pm$~0.66 & 228 & -1.5 & 
 & -11.73~$\pm$~7.28 & 229 & -1.5 \\ 

$\delta \dot{e}_{n,\ell}$....... & -0.21~$\pm$~0.94 & 229 & -0.2 & 
 & -0.21~$\pm$~0.88 & 228 & -0.2 &  
 & -1.24~$\pm$~10.13 & 229 & -0.1 \\ 

\cutinhead{$30\leq\ell\leq60$ and $1800\leq\nu\leq3600~\mu$Hz}

$\delta h_{n,\ell}$....... & -1.23~$\pm$~0.95 & 291 & -1.2 & 
 & -1.26~$\pm$~0.91 & 291 & -1.2 &  
 & -15.06~$\pm$~9.54 & 291 & -1.4 \\ 

$\delta\gamma_{n,\ell}$....... & 0.57~$\pm$~0.72 & 285 & 1.0 & 
 & 0.57~$\pm$~0.70 & 285 & 1.0 & 
 & 7.80~$\pm$~7.42 & 288 & 1.2 \\ 

$\delta p_{n,\ell}$....... & -0.71~$\pm$~0.51 & 289 & -1.1 & 
 & -0.75~$\pm$~0.47 & 289 & -1.2 & 
 & -7.81~$\pm$~5.31 & 291 & -1.1 \\ 

$\delta \dot{e}_{n,\ell}$....... & -0.19~$\pm$~0.83 & 290 & -0.2 & 
 & -0.24~$\pm$~0.80 & 288 & -0.3 & 
 & -0.50~$\pm$~8.67 & 284 & -0.1 \\ 

\cutinhead{$30\leq\ell\leq60$ and $2700\leq\nu\leq3300~\mu$Hz}

$\delta h_{n,\ell}$....... & -1.97~$\pm$~0.74 & 105 & -1.7 &
 & -1.98~$\pm$~0.69 & 105 & -1.7 & 
 & -22.76~$\pm$~7.32 & 105 & -1.8 \\ 

$\delta\gamma_{n,\ell}$....... & 1.10~$\pm$~0.53 & 105 & 1.4 & 
 & 1.07~$\pm$~0.49 & 105 & 1.5 & 
 & 13.17~$\pm$~5.39 & 105 & 1.7 \\ 

$\delta p_{n,\ell}$....... & -0.91~$\pm$~0.45 & 105 & -1.3 & 
 & -0.95~$\pm$~0.42 & 105 & -1.4 & 
 & -10.09~$\pm$~4.54 & 105 & -1.4 \\ 

$\delta \dot{e}_{n,\ell}$....... & 0.11~$\pm$~0.62 & 105 & 0.1 & 
 &  0.04~$\pm$~0.58 & 105 & 0.0 & 
 & 2.12~$\pm$~6.57 & 105 & 0.2 \\ 
\enddata 

\tablecomments{The table shows the mean fractional changes $\langle \delta Y_i \rangle$ 
               times $10^3$ per unit of change in solar activity proxy and their corresponding
               standard deviations $\sigma_{\langle \delta Y_i \rangle}$;
               the number of multiplets $N$; and $\sigma$, the mean fractional changes divided
               by the standard deviations of the distributions of 2500 randomized data sets.}
\end{deluxetable} 

\begin{deluxetable}{rrrr} 
\tablecolumns{6} 
\tablewidth{0pt} 
\tabletypesize{\footnotesize}
\tablecaption{{Global changes (\%) from minimum to maximum of the solar cycle averaged between
$30\leq\ell\leq60$ and $2700\leq\nu\leq3300~\mu$Hz}
\label{table:global}}
\tablehead{ 
\colhead{$\delta Y_i$}    &  \colhead{$F_{10.7}$} & \colhead{$R_I$} &  \colhead{KPMI}}  
\startdata 
$\delta h_{n,\ell}$....... & -23.7~$\pm$~8.9 & -26.7~$\pm$~9.3  &  -25.0~$\pm$~8.1\\

$\delta\gamma_{n,\ell}$....... & 13.2~$\pm$~6.3 & 14.5~$\pm$~6.7 & 14.5~$\pm$~5.9 \\

$\delta p_{n,\ell}$....... & -10.9~$\pm$~5.4  & -12.9~$\pm$~5.6  & -11.1~$\pm$~5.0 \\

$\delta \dot{e}_{n,\ell}$....... & 1.3~$\pm$~7.5 & 0.5~$\pm$~7.9   & 2.3~$\pm$~7.2\\
\enddata

\tablecomments{The observed changes in $F_{10.7}$, $R_I$, and KPMI indices used in the table 
are 120 ($10^{-22}\rm{W}~\rm{m}^{-2}\rm{Hz}^{-1}$), 135 (dimensionless) and 11 (G) 
respectively.}
\end{deluxetable}
\clearpage

\section{SIGNIFICANCE OF THE OBSERVED CHANGES}
\begin{figure*}[t]
\centering
\epsscale{1}
\plotone{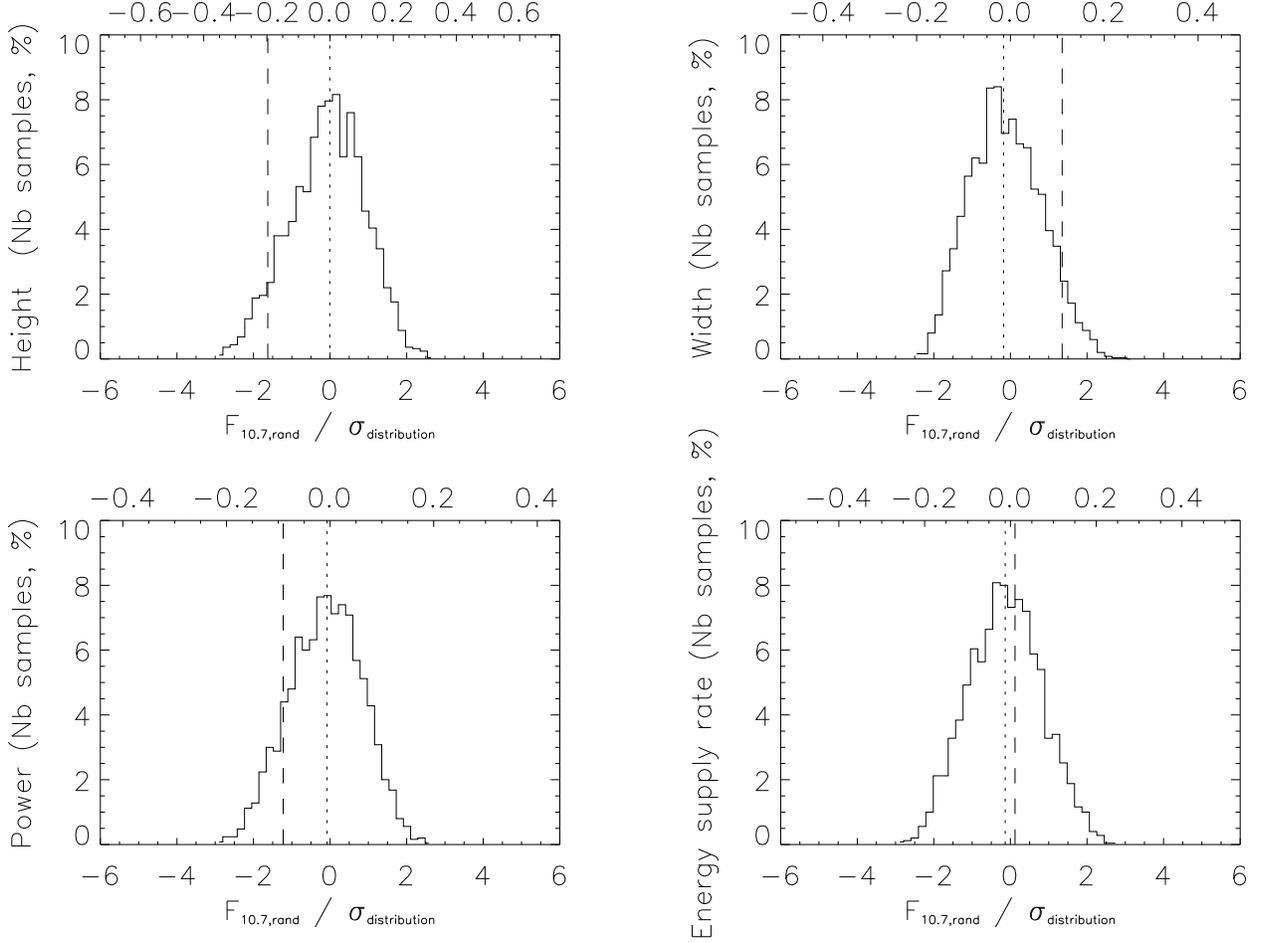}
\caption{Distributions of the fractional changes in mode height, mode width, mode velocity power
and mode energy supply rate (normalized by the standard deviations of the distributions, lower $x$-axis) obtained 
from 2500 randomized $F_{10.7}$ samples, averaged between $\ell=30$ 
and $\ell=60$ and $2700\leq\nu\leq3300~\mu$Hz. The upper $x$-axis are in the actual observed change units. 
The dotted lines correspond to the means of the distributions,
and the dashed lines to the actual observations (Figures~\ref{fig:fracnuflux}-\ref{fig:fracdegflux}).}
\label{fig:dist}
\end{figure*}

In order to check the significance of the observed changes in the p-mode parameters, 
we followed the same methodology as described in \citet{komm00a}.
We randomized in time the values of the activity proxies ($F_{10.7}$, $R_I$ and KPMI), 
and computed the regression using equation~\ref{eq:lincorr} as a function
of the fill factor and the shuffled activity index. This process was repeated 2500 
times, making sure
to not use the same randomization twice. Figure~\ref{fig:dist} shows the
distributions of the fractional changes for the mode height, 
mode width, mode velocity power and mode energy supply rate. They were 
averaged between $\ell=30$ and $\ell=60$ and  $2700\leq\nu\leq3300~\mu$Hz, and normalized by
the standard deviation of the distributions. The dotted lines
represent the corresponding means of each distribution. The dashed lines are the averages
of the real observations.

In the case of the radio flux, $F_{10.7}$, the actual average fractional 
changes in mode height and 
mode width for modes  between $\ell=30$ and $\ell=60$ in the frequency range
from 2700 to 3300~$\mu$Hz are respectively -1.7 and 1.4 standard deviations of the 
randomized distribution 
away from zero, and -1.3 for the mode power. The energy supply rate is 0.1 standard 
deviation away from zero.
The observed changes with solar activity of the mode height, mode width, and mode velocity power
are then significant to
about 90\%, whereas the changes in mode energy supply rate is not significant, which 
is then consistent with a 
zero change. Similar results over different ranges of mode degrees and frequencies 
are found, and are summarized in Table~\ref{table:change}.

\begin{figure*}[t]
\centering
\epsscale{1}
\plotone{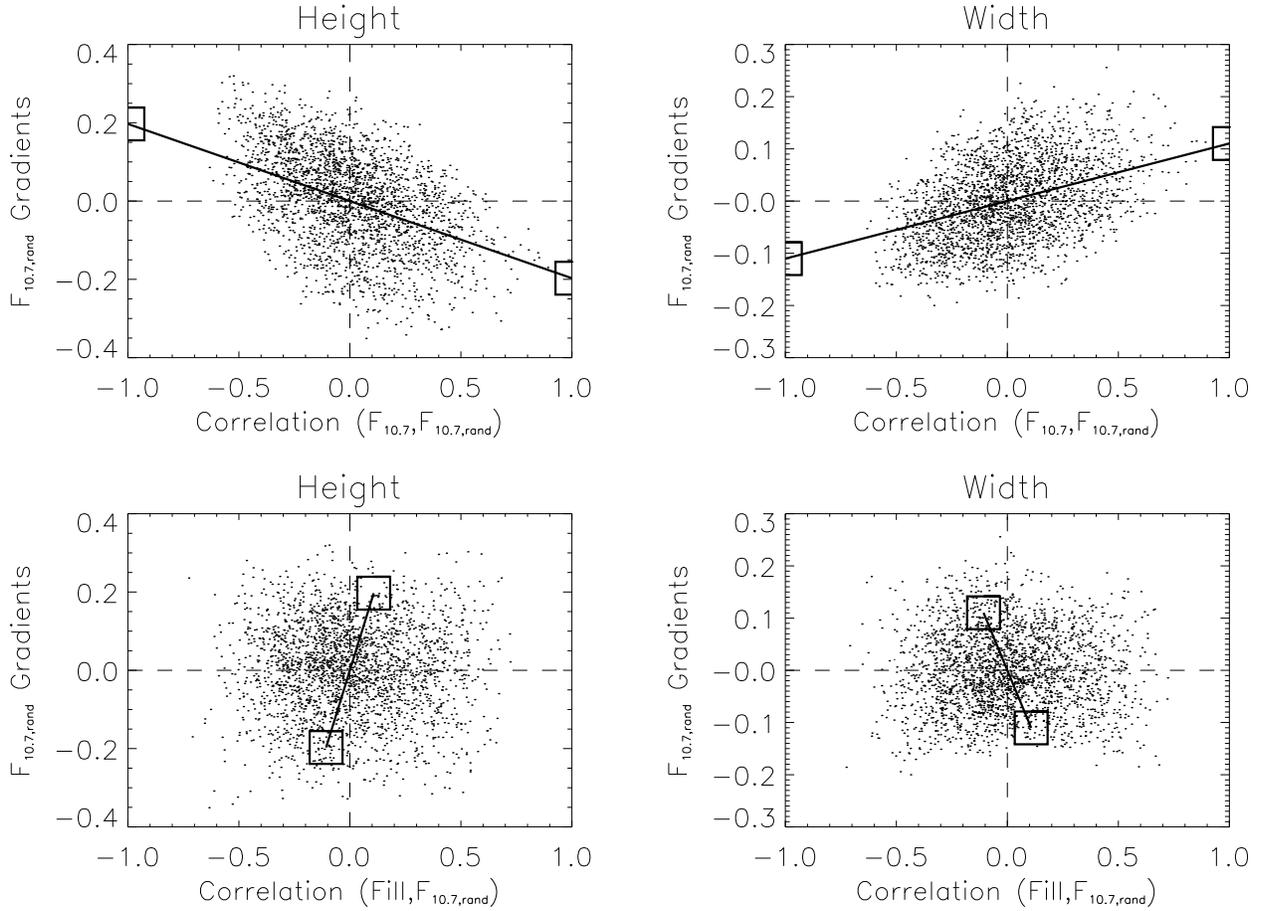}
\caption{Fractional changes in mode height and mode width
obtained from 2500 randomized $F_{10.7}$ samples as a function of 
the correlation coefficients between the measured
and randomized radio flux ({\it upper row}) and between the fill value and 
the randomized radio flux ({\it lower row}), averaged between $\ell=30$ 
and $\ell=60$ and $2700\leq\nu\leq3300~\mu$Hz. The squares connected
by a solid line correspond to the actual values.}
\label{fig:corr1}
\end{figure*}

As a second check, we also computed the correlation coefficients between 
the randomized activity indices and their corresponding
real observations \citep[see][]{komm00a}. The upper panels of Figure~\ref{fig:corr1} show the correlation 
coefficients plotted versus
the changes in the randomized sets for the mode height and mode width, 
averaged between $\ell=30$ and $\ell=60$ and from 2700 to 3300$~\mu$Hz. 
The correlation in mode height and mode width show
an elongated distribution along the direction of the actual observed changes (square) and
centered around zero. This 
elongation is present also for the average over all the mode degrees and frequencies. 
The average
randomized changes get closer to the actual change as long as the correlation between 
the randomized
activity and the measured activity increases and gets closer to 1. Similar results
 are found for the changes
in mode velocity power. As for the energy supply rate,
the correlation coefficients are a cloud of points centered around zero, and contrary to
the other parameters, do not show a prefered direction, which is consistent again 
with a zero change.

As a last test, and in order to check that the correction of the window 
function included in equation~\ref{eq:lincorr} 
does not introduce any bias in the obtained p-mode parameter changes 
with solar activity, we computed the 
correlation between the randomized activity proxies and the fill values \citep[see][]{komm00a}. 
The lower panels
on Figure~\ref{fig:corr1} present these correlation coefficients plotted versus 
the changes in the randomized sets for the mode height and mode width, averaged 
between $\ell=30$ and $\ell=60$ and from 2700 to 3300~$\mu$Hz. No prefered 
directions are observed, and 
the points are well distributed around zero, meaning that no bias is introduced 
by the window function correction. The changes in mode velocity power and energy supply rate show 
similar clouds of points.
Similar results are also obtained over different degree and frequency ranges.

We also performed the same tests with the sunspot number $R_I$ and the KPMI index, 
and similar results were found (Table~\ref{table:change}). 
With regard to these different tests, 
we can conclude that the observed changes in the p-mode damping and
excitation parameters are related to solar activity. However, as for the low-degree modes ($\ell\leq3$), 
their uncovered temporal variations were not observed to be statistically significant at the level 
of precision of the observations. Indeed, the most reliable measurements for low-degree modes come from full-disk Sun-as-a-star instruments.

In order to determine the significances of the frequency dependence reported in Section~\ref{sec:results}, 
the gradients on frequency were extracted by performing weighted linear fits on the 100~$\mu$Hz averages from Figure~\ref{fig:fracnuflux} on both sides of 3100~$\mu$Hz. The errors on the fits were used to determine the 
significance of the returned gradients for each of the studied parameters.
As a second test, we randomized the order of the frequency values and performed the same linear regression as above on both
sides of 3100~$\mu$Hz on the 100~$\mu$Hz averages. The process was repeated 250 times, making sure to not use the same randomization twice. The distributions of the returned gradients obtained from the randomized frequencies allowed us to measure the significances of the observed frequency dependences. Both methods return consistent results for the three solar
proxies. The frequency dependences of the fractional changes in mode height, mode width and mode velocity power are significant by more than 3~$\sigma$. The gradients for the mode energy supply rate are closer to 1~$\sigma$, indicating that they
are consistent with no frequency dependence at the level of precision of the data.
\clearpage

\clearpage

\section{SUMMARY AND DISCUSSION}
We used the LOWL observations collected between beginning of 1994 and 
end of 1999 to study the temporal variations of the p-mode damping and excitation parameters with solar activity 
over a wide range of angular degrees.
We extracted the temporal variations in mode height, 
mode width, mode velocity power and mode energy supply rate by
applying a multi-linear regression (eq.~\ref{eq:lincorr}) on the estimates of 16 108-d independent timeseries.
This regression allowed us to correct the bias on the peak-finding estimates from the temporal fills and 
to obtain a measure of their fractional changes.
We showed that the p-mode damping and excitation parameters between $\ell=0$ and $\ell=60$ 
present different behaviours with solar activity.
Similar results are obtained with 3 different solar activity proxies (the integrated solar radio flux, $F_{10.7}$; 
the sunspot number, $R_I$; and the Kitt Peak Magnetic Index, KPMI). Our findings
confirm the previous measurements at intermediate-$\ell$ modes obtained with GONG observations \citep{komm00a,komm00b}.

The mode heights decrease with increasing activity and show the largest variations among the
mode parameters. The p-mode damping (proportional to the mode widths) shows an increase with
an increasing activity. Just as with the mode heights, the mode velocity power
decreases with an increasing activity. Their sensitivities to solar activity show clear evidence of a frequency dependence,
the modes in the frequency range from 2700 to 3300~$\mu$Hz showing the largest variations
and presenting a maximum change centered around 3100~$\mu$Hz, as already reported in GONG observations \citep{komm00a}.
In contrast to the other parameters, the p-mode excitation does not show a significant frequency dependence
and is consistent with a zero change along the solar cycle.

As a function of the mode degree, the variations in the damping and excitation p-mode parameters are more
or less $\ell$-independent over the range of studied degrees as observed by \citet{komm00a} with GONG data. 
However, in the limit of precision of the data and as already mentioned in \citet{komm00a}, there might be some hints of a decrease in the changes in mode height, mode velocity power and mode energy supply rate for $\ell<30$, while the mode damping variations do not present an increase with decreasing $\ell$-value as suggested by \citet{komm00a}.

The relative sizes of the uncovered changes 
at intermediate $\ell$ are consistent with an increase of the damping rate over the solar cycle, 
while the net forcing of the modes remains constant. Then, in the description of the p modes by a
damped, forced oscillator, the sizes and signs they uncovered can result from changes only to the net damping, 
as already pointed out by \citet{chaplin00} for low-$\ell$ modes.

Our findings about the energy supply rate are consistent with the results obtained
with GONG data \citep{komm00b}, showing no significant changes in the p-mode excitation with solar activity, 
but we cannot rule out that the energy supply rate might change locally, since the peak-finding procedure used
in this analysis minimized only one height and one width per each ($\ell,m$) multiplet. 
However, \citet{komm02} stated that
"within the limits  of the current measurements (i.e., GONG observations), the energy supply rate 
does not sense the latitudinal distribution of magnetic activity".

\acknowledgments
This work was supported by NSF through base funding of HAO/NCAR
and partially funded by the grant AYA2004-04462 of the Spanish
National Research Plan. The authors acknowledge the LOWL observers 
Eric Yasukawa and Darryl Koon. NSO/Kitt Peak magnetic data used here 
are produced cooperatively by NSF/NOAO, NASA/GSFC and NOAA/SEL.



\clearpage

\end{document}